\documentclass[preprint]{ptephy_v1}%%%%%% to generate preprint number
\usepackage{amsmath} %for dealing with mathematics,
\usepackage{amsthm}% for dealing with theorem environments,
\usepackage{graphicx} %for dealing with figures.\usepackage{algorithmic} %for describing algorithms
\usepackage{subfig} %for getting the subfigures e.g., "Figure 1a and 1b" etc.
\usepackage{url} %It provides better support for handling and breaking URLs.
\usepackage{color}
\usepackage{bm,physics}
\usepackage{amsmath,amssymb}
\usepackage{ulem, color}
\usepackage{amsmath}
\usepackage{natbib}
\usepackage{mathrsfs}
\bmdefine{\ba}{a}
\bmdefine{\bb}{b}
\bmdefine{\bx}{x}
\bmdefine{\by}{y}
\bmdefine{\bz}{z}
\bmdefine{\bn}{n}
\bmdefine{\bp}{p}
\newcommand{\BM}{\begin{pmatrix}}
\newcommand{\EM}{\end{pmatrix}}

\newif\ifHIDEHIGHLIGNT
\HIDEHIGHLIGNTtrue %%%%% highlight the alterations
\ifHIDEHIGHLIGNT
\usepackage{ulem,color}
\newcounter{nnn}

\else

\fi
\begin{document}
\title {
Equivalence of generator coordinate  Brink  cluster model  and      nonlocalized  cluster model and  
  supersolidity of
      $\alpha$ cluster structure  in nuclei
}

%
%%%% To generate auto affiliation numbers please use \author{}\affil{} command
\author{S. Ohkubo}

\affil{Research Center for Nuclear Physics, Osaka University,
Ibaraki, Osaka 567-0047, Japan }
%\email{ohkubo@rcnp.osaka-u.ac.jp}

\begin{abstract}
It is found that  $\alpha$ cluster structure 
has the apparently opposing dual property of crystallinity and condensation simultaneously.
 The
 mathematical equivalence of the spatially localized Brink  $\alpha$ cluster model in the  generator coordinate method (GCM)
   and the nonlocalized cluster model (NCM), which is also called the THSR (Tohsaki-Horiuchi-Schuck-R$\ddot{\rm o}$pke) wave function based on the condensation of $\alpha$ clusters, is  shown.  The latter is found to be  an equivalent representation of the localized cluster model and it is a natural consequence that the many NCM (THSR) calculations reproduce the proceeding cluster model calculations using  the  GCM and the resonating group method (RGM).  Localized  cluster models, which have been successfully used  for  more than half a century, will  continue to be very powerful. The equivalence is a manifestation of the duality of incompatible   aspects: crystallinity and   coherent wave nature due to condensation of  $\alpha$ clusters, i.e.  the dual  properties  of a supersolid.
 The Pauli principle   causes the duality.
  The evidence for supersolidity, the emergence of a Nambu-Goldstone  mode caused by the spontaneous symmetry breaking of the global phase, is discussed. 
   \end{abstract}

\subjectindex{xxxx, xxx}
%\pacs{21.60.Gx,27.20.+n,67.85.De,03.75.Kk}
\maketitle

\par
The $\alpha$ particle, which is  the most tightly bound 
nucleus,
 has been playing an important role 
  for more than a century in  quantum physics and nuclear physics. The observation of  anomalously large backward angle  scattering of $\alpha$ particles from a gold target lead Rutherford to the discovery of the nucleus  \cite{Rutherford1911}.  The $\alpha$ particle emission from radioactive  nuclei was understood by Gamow  \cite{Gamow1928} and 
Gurney and  Condon \cite{Gurney1928,Gurney1929} for the first time as quantum  tunneling, 
which  has been  shed light  recently from the viewpoint  of  Lefschetz thimble \cite{Tanizaki2014}.
The $\alpha$ particle model, in which  the $\alpha$ particle is considered as a constituent unit of the nucleus, was proposed as the first nuclear structure model in 1937 \cite{Wefelmeier1937,Wheeler1937,Wheeler1937B}.  While  the naive classical $\alpha$ particle model was criticized
  \cite{Blatt1952}
 in the advent of  the shell model \cite{Mayer1948,Haxel1949} and the collective model \cite{Bohr1952}, the successful  shell model and the collective model  \cite{Bohr1969A,Bohr1969B,Ring1980} also encountered   difficulty  explaining the emergence of  very low-lying  intruder  states in light  nuclei such as the mysterious  $0_2^+$ (6.05 MeV) state in $^{16}$O \cite{Arima1967,Marumori1968}. The developed  quantum   $\alpha$ cluster model, in which the effect of the Pauli principle is taken into account, has witnessed  tremendous success in recent decades  in explaining both shell-model like states and  $\alpha$ cluster states comprehensively, which are reviewed  in  light nuclei in Refs.   \cite{Suppl1972,Wildermuth1977,Suppl1980} and in the medium-weight nuclei
  in Ref. \cite{Suppl1998}.
 
\par
 Cluster models such as the Brink cluster model using the  generator coordinate method (GCM) \cite{Brink1966}, the resonating group method (RGM), which is equivalent to the GCM \cite{Horiuchi1977},  the orthogonality condition model (OCM) \cite{Saito1969},  and the local potential model (LPM) \cite{Buck1975,Ohkubo1977,Michel1983,Ohkubo2016}, are all based  on the picture that  the  cluster structure   has  a geometrical configuration. Examples are the  two $\alpha$ dumbbell structure of $^8$Be \cite{Horiuchi1970,Hiura1972}, the three $\alpha$ triangle structure of  $^{12}$C \cite{Uegaki1977,Uegaki1979,Kamimura1977}, and the $\alpha$+$^{16}$O structure in $^{20}$Ne \cite{Horiuchi1972,Hiura1972B,Fujiwara1980}.
      In recent decades structure studies     using    cluster models in the bound and quasi-bound energies  have all supported  the spatially localized cluster picture \cite{Suppl1972,Wildermuth1977,Suppl1980,Suppl1998}.  Also  the unified understanding of  cluster structure in the  low energy region and 
    prerainbows  and  nuclear rainbows in the scattering region, which are   confirmed for  the systems such as $\alpha$+$^{16}$O and $\alpha$+$^{40}$Ca  \cite{Michel1998,Ohkubo2016}, supports   the geometrical localized cluster picture.

\par % nonlocalized
  On the other hand,   calculations using the  nonlocalized cluster model (NCM), 
   which was originally proposed to explain the dilute gas-like $0_2^+$ Hoyle state in $^{12}$C from the viewpoint of Bose-Einstein condensation (BEC) of  $\alpha$ clusters  \cite{Tohsaki2001} and named   THSR wave function for  the authors
(Tohsaki-Horiuchi-Schuck-R$\ddot{\rm o}$pke)     \cite{Funaki2009},
      reproduced 
       the $\alpha$ cluster structure in  $^8$Be, $^{12}$C and $^{20}$Ne   \cite{Funaki2002,Funaki2003,Funaki2005,Zhou2012,Zhou2013,Zhou2014} almost as well as  the preceding GCM and RGM calculations.
In contrary to the traditional  geometrical localized  cluster picture, the NCM  calculations  conclude that the most typical nucleus, $^8$Be,
      is more of a very dilute gas-like structure of two $\alpha$ cluster  rather than  a  solid dumbbell structure of two $\alpha$ clusters \cite{Funaki2002}.
       The two concepts of gas and solid are opposing.
Furthermore because the  two pictures are based on the 
 incompatible concepts, from the viewpoint of the NCM (THSR) it 
 has been summarized in Ref. \cite{Zhou2014}  that the traditional
understanding is incorrect, namely, that the localized
  cluster picture is strongly supported by the energy curve with the Brink  wave  function which gives the minimum point at a nonzero
value of the intercluster distance parameter.
    
     Why the NCM calculations   based on the apparently  exclusive concept give  similar results to the preceding  GCM and RGM calculations based on the localized cluster  picture has not been understood and  has remained puzzling for the  last 
       two decades.  Very recently Ohkubo {\it et al.} used a superfluid $\alpha$ cluster model  to  report \cite{Ohkubo2020} that this 
      puzzle  can  be solved by noticing that the $\alpha$ cluster structure has a duality of crystallinity (localization) and condensation (nonlocalization),
      a property of supersolidity. 
According to this theory,   while the former is the view  from the particle nature of the cluster structure, the latter is the view from  the wave nature due to the coherence of the condensate cluster structure and the two are compatible.  
        It is important to reveal generally and rigorously the relation between the          geometrical localized  cluster picture and the nonlocalized  cluster picture %based on 
         and to deepen the underlying physical meaning of the $\alpha$ cluster structure.

\par
In  this paper it is  shown that the NCM, namely the  THSR  wave function,  is mathematically rigorously equivalent to the traditionally used  geometrical Brink cluster model  in the generator coordinate method.
The reason why 
  the  THSR  calculations
\cite{Funaki2002,Funaki2003,Funaki2005,Funaki2009,Zhou2012,Zhou2013,Zhou2014,Funaki2018,Zhou2019,Zhou2020}  give very similar results to   the preceding cluster model calculations using the GCM and RGM based on the geometrical picture is clarified.  
   It is shown generally  that 
    $\alpha$ cluster structure   has the duality of apparently  exclusive properties  of crystallinity (localization) and condensation (nonlocalization), i.e. supersolidity.

\par
 The $n$-$\alpha$ cluster model based on the  geometrical crystalline picture such as the two $\alpha$ cluster model of $^8$Be and the  three $\alpha$ cluster model of $^{12}$C,  is given by the following Brink wave function \cite{Brink1966}
  \begin{align}
& \Phi^{B}_{n\alpha}(\boldsymbol{R}_1, \cdots , \boldsymbol{R}_n) =\frac{1}{\sqrt{(4n)!} }{\rm det} [\phi_{0s}(\boldsymbol{r}_1- \boldsymbol{R}_1)\chi_{\tau_1,\sigma_1}]  \cdots \phi_{0s}(\boldsymbol{r}_{4n}-  \boldsymbol{R}_n)\chi_{\tau_{4n},\sigma_{4n}},
 \label{Brinkwf}
 \end{align}
  \noindent where   $\boldsymbol{R_i}$
  is a parameter that specifies the center  of the $i$-th $\alpha$ cluster. 
 $\phi_{0s} ( \boldsymbol{r} -  \boldsymbol{R})$ is a
0s harmonic  oscillator wave function with  a size parameter $b$  around a center  $\boldsymbol{R}$,
 \begin{align}
 \phi_{0s}(\boldsymbol{r} - \boldsymbol{R}) = \left( 
 \frac{1}{\pi b^2}\right)^{3/4} 
 \exp \left[  - \frac{ (\boldsymbol{r} - \boldsymbol{R})^2}{2b^2} \right], 
 \end{align}
 \noindent and  $\chi_{\sigma, \tau}$ is the spin-isospin wave function of a nucleon.
 Eq.(\ref{Brinkwf}) is rewritten  as
  \begin{align}
 & \Phi^{B}_{n\alpha}(\boldsymbol{R}_1, \cdots , \boldsymbol{R}_n)  =\mathscr{A} \left[   \prod_{i=1}^n \exp \left\{   - 2 \frac{ ( \boldsymbol{X}_i - \boldsymbol{R}_i )^2}{b^2}  \right\} \phi(\alpha_i)
  \right],
\label{Brinkwf2}
\end{align}
 \noindent where ${ \boldsymbol X_i}$ is the center-of-mass coordinate  of the $i$-th $\alpha$ cluster and   $\phi(\alpha_i$) represents the internal wave function of the $i$-th $\alpha$ cluster. $\mathscr{A}$ is the antisymmerization operator.
 The generator coordinate wave function  $\Psi_{n\alpha}^{GCM}$ based on the geometrical configuration of the Brink wave function is  given by
 \begin{align}
& \Psi_{n\alpha}^{GCM}=\int d^3  \boldsymbol{R}_1 \cdots d^3 \boldsymbol{R}_n f( \boldsymbol{R}_1, \cdots , \boldsymbol{R}_n ) 
\Phi^{B}_{n\alpha}( \boldsymbol{R}_1, \cdots ,  \boldsymbol{R}_n).  
  \label{GCMwf1}
\end{align} 

\par
We show that the localized cluster model of Eq.~(\ref{GCMwf1}) and the nonlocalized cluster model 
are mathematically equivalent. For the sake of simplicity we  treat  hereafter
  the simplest two $\alpha$ cluster structure of  $^8$Be.
The generator  coordinates  $\boldsymbol{R}_1$ and  $\boldsymbol{R}_2$, which specify the position parameters of the two  $\alpha$ clusters, are rewritten   as follows by using $\boldsymbol{R}_G$ and $\boldsymbol{R}$, which are the center-of-mass  and  the relative vectors, respectively,
\begin{align} 
    \boldsymbol{R}_1 = \boldsymbol{R}_G +  \frac{1}{2} \boldsymbol{R}, \quad  
    \boldsymbol{R}_2 = \boldsymbol{R}_G -  \frac{1}{2} \boldsymbol{R}.
  \label{com}
\end{align}  
We take $ \boldsymbol{R}_G$=0 to remove the spurious center-of-mass motion and use  the notation $\Phi^{B}_{2\alpha}( \boldsymbol{R})$  
 for $ \Phi^{B}_{2\alpha}( \boldsymbol{ \frac{1}{2}R},   \boldsymbol{ -\frac{1}{2}R})$.  Thus  Eq.~(\ref{GCMwf1}) is written as
\begin{align} 
&  \Psi_{2\alpha}^{GCM}=\int d^3   \boldsymbol{R} f(\boldsymbol {R}) 
\Phi^{B}_{2\alpha}(  \boldsymbol{R}).
  \label{GCMwf8Be}
 \end{align} 
\noindent   We introduce ${g}(\boldsymbol{\mu})$, which is related to $f(\boldsymbol{R})$  by the Laplace transformation
 \begin{align} 
&    f(\boldsymbol {R}) =\int_{0}^{\infty} d\mu_x  \int_{0}^{\infty} d\mu_y \int_{0}^{\infty} d\mu_z\exp\left[-(\mu_x {R_x}^2+\mu_y {R_y}^2+\mu_z {R_z}^2)\right] {g}( \boldsymbol{\mu}), 
\label{Laplace}
\end{align} 
\noindent where    $\boldsymbol{\mu}=(\mu_x, \mu_y, \mu_z)$.
Then Eq.(\ref{GCMwf8Be})  reads 
\begin{align} 
&  \Psi_{2\alpha}^{GCM}=\int d^3   \boldsymbol{\mu} {g}(\boldsymbol {\mu}) \left[\int d^3  \boldsymbol{R}\exp \left\{-(\mu_x {R_x}^2+\mu_y {R_y}^2+\mu_z {R_z}^2)\right\}
\Phi^{B}_{2\alpha}(\boldsymbol{R})\right].
  \label{GCMwf8Be2}
 \end{align} 
\noindent The  term  $\left[\cdots\right]$ in the rhs of Eq.(\ref{GCMwf8Be2}) is nothing but the  definition of the NCM (THSR) wave function $\Phi_{2\alpha}^{NCM}$
    for the two $\alpha$ clusters of $^8$Be \cite{Zhou2012,Zhou2013,Zhou2014},
  \begin{align} 
&  {\Phi}_{2\alpha}^{NCM}(\boldsymbol{\mu}) \equiv 
 \int d^3  \boldsymbol{R}\exp \left\{-(\mu_x {R_x}^2+\mu_y {R_y}^2+\mu_z {R_z}^2)\right\}\Phi^{B}_{2\alpha}(\boldsymbol{R}),
  \label{NCMwf8Be2-1}
  \end{align} 
 \begin{align}
& \qquad \quad 
\propto 
 \mathscr{A} \left[     \prod_{i=1}^{2}  \exp \left\{  - 2 \left(\frac{ X_{ix}^2}{B_{x}^2}+\frac{ X_{iy}^2}{B_{y}^2}+\frac{ X_{iz}^2}{B_{z}^2} \right) \right\}  \phi(\alpha_i)       \right],
\label{8BeNCM2}
\end{align}
\noindent where
\begin{eqnarray}
& B_{k} =\sqrt{b^2 + {\mu_{k}}^{-1} } \quad   (k=x, y, z).
 \label{SizeParameter}
\end{eqnarray}
\noindent Eq.(\ref{8BeNCM2})  is the internal wave function and is independent of the center-of-mass motion.
 Eq.(\ref{GCMwf8Be2})   reads 
\begin{align}
&  \Psi_{2\alpha}^{GCM}=\int d^3   \boldsymbol{\mu} {g}( \boldsymbol{\mu})  {\Phi}_{2\alpha}^{NCM}(\boldsymbol{\mu}).
 \label{GCMNCMwf8Be2-1}
  \end{align}  
  
\noindent  While  in Eq.(\ref{GCMwf8Be}) the GCM wave function is expressed  based on the geometrical picture using the Brink function $\Phi^{B}_{2\alpha}(\boldsymbol{R})$ as a base function, in  Eq.(\ref{GCMNCMwf8Be2-1})   the same GCM wave function is expressed   using the nonlocalized wave function ${\Phi}_{2\alpha}^{NCM}$ as a base function.

\par 
 The weight function $f( \boldsymbol{R})$ in Eq.(\ref{GCMwf8Be}) is determined
by the variation principle 
 \begin{align} 
\delta \frac{ < \Psi_{n\alpha}^{GCM}|H| \Psi_{n\alpha}^{GCM}>}{< \Psi_{n\alpha}^{GCM}| \Psi_{n\alpha}^{GCM}>}=0. 
  \label{variation}
\end{align} 
This leads to
   the Hill-Wheeler equation,
  \begin{align} 
 \int d^3  \boldsymbol{R}^{\prime} [  \mathscr{H}(\boldsymbol{R}, \boldsymbol{R}^{\prime})-E  \mathscr{B}(\boldsymbol{R}, \boldsymbol{R}^{\prime})] f(\boldsymbol{R}^{\prime})  =  0, 
  \label{HillWheeler1}
\end{align} 
\noindent where
  \begin{align} 
   \mathscr{H}(\boldsymbol{R}, \boldsymbol{R}^{\prime})  =  < \Phi_{n\alpha}^{B}(\boldsymbol{R})|H| \Phi_{n\alpha}^{B}( \boldsymbol{R}^{\prime})>,
  \label{HillWheeler2}
\end{align} 
\noindent and 
  \begin{align} 
   \mathscr{B}(\boldsymbol{R}, \boldsymbol{R}^{\prime})  =  < \Phi_{n\alpha}^{B}(\boldsymbol{R})| \Phi_{n\alpha}^{B}( \boldsymbol{R}^{\prime})>.
     \label{HillWheeler3}
\end{align}  
\noindent  $H$ and $E$  are
  the Hamiltonian of the system and the eigenenergy, respectively. 

\par
Similarly   the weight function $g( \boldsymbol{\mu})$  in Eq.(\ref{GCMNCMwf8Be2-1}) is determined  by solving the following 
 Hill-Wheeler equation for  $g( \boldsymbol{\mu})$,
  \begin{align} 
 \int d^3  \boldsymbol{\mu}^{\prime} [  \mathscr{H}^{NCM}(\boldsymbol{\mu}, \boldsymbol{\mu}^{\prime})-E  \mathscr{B}^{NCM}(\boldsymbol{\mu}, \boldsymbol{\mu}^{\prime})] g(\boldsymbol{\mu}^{\prime})  =  0, 
  \label{HillWheeler1NCM}
\end{align} 
\noindent where 
  \begin{align} 
   \mathscr{H}^{NCM}(\boldsymbol{\mu}, \boldsymbol{\mu}^{\prime})  =  < \Phi_{n\alpha}^{NCM}(\boldsymbol{\mu})|H| \Phi_{n\alpha}^{NCM}( \boldsymbol{\mu}^{\prime})>,
  \label{HillWheeler2NCM}
\end{align} 
\noindent and 
  \begin{align} 
   \mathscr{B}^{NCM}(\boldsymbol{\mu}, \boldsymbol{\mu}^{\prime})  =  < \Phi_{n\alpha}^{NCM}(\boldsymbol{\mu})| \Phi_{n\alpha}^{NCM}( \boldsymbol{\mu}^{\prime})>.
  \label{HillWheeler3NCM}
\end{align}  

 \par  We show  that  the weight function  ${g}$($\boldsymbol{\mu}$) is a $\delta$ function.
   We expand $f(\boldsymbol{R})$ using a Gaussian function as follows,
 \begin{align}
&   f(\boldsymbol{R})= \sum_{i=0}^{\infty} \sum_{j=0}^{\infty} \sum_{k=0}^{\infty}  c_{i}^{(x)} c_{j}^{(y)} c_{k}^{(z)}\exp[ -(a_{ix} R_x^2 + a_{jy} R_y^2 +a_{kz} R_z^2)],
 \label{GaussExpansion}
  \end{align}   
\noindent  where   $ c_{i}^{(x)}$,  $c_{j}^{(y)}$ and  $c_{k}^{(z)}$ are the coefficients expanded by the  Gaussian function with the width parameters $a_{ix}$, $a_{jy}$ and $a_{kz}$.
 Putting Eq.(\ref{GaussExpansion}) into Eq.(\ref{Laplace}) of   the Laplace transformation, we get
  \begin{align}
&   \sum_{i=0}^{\infty} \sum_{j=0}^{\infty} \sum_{k=0}^{\infty} c_{i}^{(x)} c_{j}^{(y)} c_{k}^{(z)} \exp[ -(a_{ix} R_x^2 + a_{jy} R_y^2 +a_{kz} R_z^2)]  \nonumber \\
& =\int_{0}^{\infty} d\mu_x  \int_{0}^{\infty} d\mu_y \int_{0}^{\infty} d\mu_z\exp\left[-(\mu_x {R_x}^2+\mu_y {R_y}^2+\mu_z {R_z}^2)\right]  g_x(\mu_x)g_y(\mu_y)g_x(\mu_z).
 \label{ExpansionRelation}
   \end{align}   
  The  $x$ component reads
  \begin{align}
&   \sum_{i=0}^{\infty} c_{i}^{(x)}  \exp[ - a_{ix} R_x^2 ] 
 =\int_{0}^{\infty} d\mu_x  \exp\left[- \mu_x {R_x}^2\right]  g_x(\mu_x).
 \label{ExpansionXcomponent}
   \end{align}   
 By using the following Laplace transformation formula with  $s=R^2_x$, $\tau=a_{ix}$ and $t=\mu_x$,
  \begin{align}
&     e^{ - \tau s} 
 =\int_{0}^{\infty}  e^{-st}\delta(t-\tau) d t ,
 \label{LaplaceFormula}
   \end{align}   
we find
   \begin{align}
&   c_{i}^{(x)}=1 \quad {\rm and} \quad  g_x(\mu_x)=   \delta(\mu_x- a_{ix}) 
\quad  {\rm for} \quad  i=i_0, \nonumber \\
&   c_{i}^{(x)}=0 \quad  {\rm for} \quad  i\ne i_0.
 \label{g(x)Formula}
   \end{align}  
Putting $c_{ijk} \equiv  c_{i}^{(x)} c_{j}^{(y)} c_{k}^{(z)}$,   Eq.(\ref{ExpansionRelation}) holds only when  
   \begin{align}
&  c_{ijk}= 1\quad  {\rm and} \quad  {g}(\boldsymbol{\mu})=\delta(\mu_x -a_{{i_0}x})  \delta(\mu_y -a_{ {j_0}y})\delta(\mu_z - a_{{k_0}z}),  \nonumber\\
&\qquad \qquad\qquad \qquad\quad
 \quad\qquad  \quad {\rm for} \quad i=i_0, j=j_0, k=k_0, \nonumber\\
&   c_{ijk}= 0  \qquad \qquad\qquad    \qquad \quad \quad  {\rm for}  \quad i\ne i_0, j\ne j_0, k\ne k_0. 
 \label{Keisu}
  \end{align}   
    Thus it is found that the weight functions, $f(\boldsymbol{R})$ in Eq.(\ref{GCMwf8Be}) and $g(\boldsymbol{\mu})$  in Eq.({\ref{GCMNCMwf8Be2-1}),
   have the   simple  functional forms
     as follows, 
  \begin{align} 
  &   f(\boldsymbol{R}) =  \exp \left[- (\alpha_x R_x^2 +\alpha_y R_y^2+\alpha_z R_z^2)\right], 
     \label{WeightFunction-f-g1} \\
& {g}(\boldsymbol{\mu})=  \delta(\boldsymbol{\mu} -\boldsymbol{\alpha}),   
\label{WeightFunction-f-g2}
  \end{align}    
  \noindent where $\boldsymbol{\alpha}$=$(\alpha_{x}, \alpha_{y},\alpha_{ z})$ is defined as
    $(\alpha_{x}, \alpha_{y},\alpha_{ z}) \equiv (a_{{i_0}x}, a_{{j_0}y},a_{ {k_0}z})$. 
    Putting Eq.(\ref{Keisu}) into Eq.(\ref{GCMNCMwf8Be2-1}), we  get
    \begin{align}
&  \Psi_{2\alpha}^{GCM}=\int d^3   \boldsymbol{\mu} \delta( \boldsymbol{\mu}-\boldsymbol{\alpha})  {\Phi}_{2\alpha}^{NCM}
(\boldsymbol{\mu}),\nonumber\\
&  = {\Phi}_{2\alpha}^{NCM}
(\boldsymbol{\alpha}).
 \label{GCMNCM0}
  \end{align}  
   \noindent  Eq.(\ref{GCMNCM0}) means  that   $\Psi_{2\alpha}^{GCM}$ is completely equivalent to a  single NCM (THSR) wave function   $\Psi_{2\alpha}^{NCM}(\boldsymbol{\alpha})$. 
  This means that the cluster wave function Eq.(\ref{GCMwf8Be}) based on the geometrical $\alpha$ cluster picture
  can be  always represented by the single  NCM wave function.

  \par
  The above discussion for the simplest two $\alpha$ cluster system 
 can be    generalized to the $n$-$\alpha$ cluster system.  
  The Laplace transformation relation 
   is   generalized to  
   \begin{align} 
&    f( \boldsymbol{R}_1, \cdots , \boldsymbol{R}_n) =
\int_{0}^{\infty} 
   d\boldsymbol{\mu}
\exp\left[- \sum_{i=1}^{n} (\mu_x R_{ix}^2+\mu_y R_{iy}^2+\mu_z R_{iz}^2)\right] {g}( \boldsymbol{\mu}). 
\label{Laplace2}
\end{align} 
The nonlocalized cluster model wave function for $n$-$\alpha$ clusters is given by 
\begin{align} 	
& \Phi_{n\alpha}^{NCM}( \boldsymbol{\mu})=\int d^3  \boldsymbol{R}_1 \cdots d^3 \boldsymbol{R}_n  \exp \left[  -\sum_{i=1}^{n}  (\mu_{x} R_{ix}^2+\mu_{y} R _{iy}^2 + \mu_{z} R_{iz}^2)    \right] 
\Phi^{B}_{n\alpha}( \boldsymbol{R}_1, \cdots ,  \boldsymbol{R}_n),   
\label{NCM1}
\end{align}
 \begin{align}
& \propto  \mathscr{A} \left[     \prod_{i=1}^{n}  \exp \left\{  - 2 \left(\frac{ X_{ix}^2}{B_{x}^2}+\frac{ X_{iy}^2}{B_{y}^2}+\frac{ X_{iz}^2}{B_{z}^2} \right) \right\} 
 \phi(\alpha_i)  
     \right].
\label{NCM2}
\end{align}
Then Eq.(\ref{GCMwf1}) reads
\begin{align}
&  \Psi_{n\alpha}^{GCM}=\int d^3   \boldsymbol{\mu} {g}( \boldsymbol{\mu})  {\Phi}_{n\alpha}^{NCM}(\boldsymbol{\mu}).
 \label{GCMNCMwf-Nalpha}
  \end{align}  
Similar to the two $\alpha$ cluster case,   ${g}( \boldsymbol{\mu})$ is shown to be a $\delta$ function. Then
\begin{align}
 &  \Psi_{n\alpha}^{GCM}=\int d^3   \boldsymbol{\mu} 
	\delta( \boldsymbol{\mu}-\boldsymbol{\alpha})  	{\Phi}_{n\alpha}^{NCM}(\boldsymbol{\mu}),
\nonumber  \\
&  = {\Phi}_{n\alpha}^{NCM}(\boldsymbol{\alpha}).
	\label{GCM=NCM-Nalpha-2}
\end{align}

\par
Eq.(\ref{GCM=NCM-Nalpha-2}) means  $\Psi_{n\alpha}^{GCM (J \pi)}$=   ${\Phi}_{n\alpha}^{NCM (J \pi)}$ for any $J^\pi$ states where $\Psi_{n\alpha}^{GCM (J \pi)}$ =
$\hat{P}^{\pi}$$ \hat{P}^{J}_{MK}$$\Psi_{n\alpha}^{GCM}$  and    ${\Phi}_{n\alpha}^{NCM (J \pi)}$=$\hat{P}^{\pi}$$ \hat{P}^{J}_{MK}$${\Phi}_{n\alpha}^{NCM}(\boldsymbol{\alpha})$    with    $\hat {P}^{\pi}$ and    $\hat{P}^{J}_{MK} $ being the parity projection  and  the angular momentum projection operators, respectively.
Therefore the numerically calculated squared overlap, 
$\kappa$=$|<$ $\Psi_{n\alpha}^{GCM (J \pi)}$$|$ ${\Phi}_{n\alpha}^{NCM (J \pi)}$$>|^2$,
should be  unity for any $J^\pi$ states if the computations 
 of   $\Psi_{n\alpha}^{GCM (J \pi)}$ and  ${\Phi}_{n\alpha}^{NCM (J \pi)}$ are  accurate enough.
  For the two $\alpha$ cluster calculations  of $^8$Be
  \cite{Funaki2002}, 
   $\kappa$=0.9980 for the ground state $0^+$. 
For the three $\alpha$ cluster  calculations  of $^{12}$C, 
$\kappa$=0.93 for $0_1^+$, $\kappa$=0.90 for $2_1^+$ \cite{Funaki2018}, $\kappa$=  0.95-0.97
    \cite{Funaki2003,Funaki2005} and 0.99 
    \cite{Funaki2018} for $0_2^+$, 
   $\kappa$=0.96 for  $3_1^-$   \cite{Zhou2019} and  $\kappa=$0.92  for  $4_1^-$   \cite{Zhou2019}.
   For the  
  the   three $\alpha$ linear  chain structure 
     $\kappa$= 0.987 ($0^+$),  0.989 ($2^+$) and 0.981  ($4^+$) \cite{Suhara2014}.
For the four $\alpha$ cluster calculations of $^{16}$O 
\cite{Funaki2010,Funaki2018}, $\kappa$=0.98,  0.98, 0.98 and 0.96 for  $0^+$ (g.s.),  $0^+$ (6.05 MeV), $0^+$ (13.6 MeV) and    $0^+$ (15.1 MeV), respectively.
For the 
 four $\alpha$ linear  chain structure of $^{16}$O, 
$\kappa$=0.944 ($0^+$), 0.942 ($2^+$) and 0.931 ($4^+$)  \cite{Suhara2014}.  
 As for the  five $\alpha$  cluster structure of $^{20}$Ne, 
    $\kappa$=0.9929, 0.9879 and  0.9775 for the  ground band $0^+$,  $2^+$ and   $4^+$ states, respectively  
  \cite{Zhou2012,Zhou2013}, and $\kappa$=0.9998 and  0.9987 for the $1^-$ and   $3^-$ states of the $K=0_1^-$  band, respectively  
   \cite{Zhou2013,Zhou2014}, were reported by using the $\alpha$+$^{16}$O cluster model, which  corresponds to $f( \boldsymbol{R}_1, \boldsymbol{R}_2,\boldsymbol{R}_2,\boldsymbol{R}_2,\boldsymbol{R}_2)$
in Eq.(\ref{Laplace2}),  namely  a limiting case that the four $\alpha$ clusters  
approach the same  parameter position $\boldsymbol{R}_2=\boldsymbol{R}_3=\boldsymbol{R}_4=\boldsymbol{R}_5$  to form 
   the $^{16}$O shell model wave function centered at   $\boldsymbol{R}_2$.
 That the calculated  squared overlaps  of the single NCM wave function  with the GCM wave function
   give the values, 
 $\kappa$$\approx1$,	for all the cases reported
  is the natural consequence of the  equality of  Eq.(\ref{GCM=NCM-Nalpha-2}). The 
   calculated values   should be  $\kappa$=1    for any $J^\pi$ states described by the GCM  in more precise numerical computations.  

  \par
    Eq.(\ref{GCM=NCM-Nalpha-2}) tells us that all the physical quantities such as the energy levels, reduced widths, electric transition probabilities, root mean square radii,  {\it etc.} calculated
    using the  NCM (THSR)  wave functions of 
    Eq.(\ref{NCM1}) and Eq.(\ref{NCM2})
    are identical to those calculated using  the Brink localized cluster model in the  GCM.  Because of the  equivalence   of the GCM to  
  the RGM, it is  a natural consequence that the  numerical calculations  using the NCM
    give  almost exactly the  same wave functions to those of the  preceding RGM calculations based on the localized cluster picture. 
 It is  naturally expected  that the NCM calculations, if the computer power allows, would  give  similar physical results in  medium-weight  nuclei such as $^{40}$Ca and $^{44}$Ti  where the    OCM and LPM calculations based on the localized cluster picture have been  successful \cite{Suppl1998,Ohkubo1998}. 
    
\par   
 We consider that there should  be profound underlying physical meaning    behind  the  equivalence of the GCM wave function and its NCM representation.
Firstly, there is no doubt that the $\alpha$ cluster structure has a geometrical crystalline structure as has been evidenced by a number of theoretical and experimental studies  \cite{Suppl1972,Suppl1980,Suppl1998} of the structure in the  bound and quasi-bound energy region,  molecular resonances, ALAS (anomalous large angle scattering) or BAA (backward angle anomaly), prerainbows and nuclear rainbows in the  scattering energy region. 
For the bound and quasi-bound state energy region, that  the GCM wave function shows a  geometrical localized  $\alpha$ cluster can be seen quantitatively and intuitively  in  the GCM energy surface $V^{GCM (J,\pi)}(R)=\mathscr{H}^{J \pi}(R,R)$= $<\Phi^{B (J\pi)}_{n \alpha}({R})|H| \Phi^{B (J\pi)}_{n \alpha}({R})>$. For  the two $\alpha$ cluster structure of $^{8}$Be, the GCM   energy surface (Fig.~1 of \cite{Horiuchi1970}) shows the minimum at  $R\ne0$ ($\approx3.5$ fm), 
which  corresponds approximately to the relative distance
  in the coordinate  space between the two $\alpha$  clusters.
  This shows clearly that $^8$Be has  a dumbbell structure  of the two $\alpha$ clusters.
For the three $\alpha$ clusters, the GCM energy surfaces  (Fig.~2 of \cite{Uegaki1977})
 show the minimum   at $R\ne0$ favoring a geometrical configuration,
  equilateral triangle for the ground state $0^+$  and the $3_1^-$ state in $^{12}$C.
 Similarly the  GCM energy surfaces calculated using the $\alpha$+$^{16}$O Brink cluster model (Fig.~17 of \cite{Hiura1972B})
  show the  energy minimum at $R\ne0$  indicating a geometrical  cluster structure for the  parity-doublet $K=0_1^+$ and $K=0_1^-$ bands states in $^{20}$Ne.
In addition, 
 the existence  of  the  higher nodal band states in $^{20}$Ne, in which the intercluster relative motion is excited and  
whose   higher spin  member states
are responsible for  the
  ALAS (BAA) phenomena  in $\alpha$ scattering from $^{16}$O \cite{Ohkubo1977,Michel1983},
give  strong support to  the geometrical $\alpha$ cluster viewpoint. 
The existence of 
 a  higher nodal band   with  $\alpha$ cluster structure has been   also  confirmed in  medium-weight  nuclei such as 
   $^{40}$Ca \cite{Yamaya1994,Sakuda1994,Sakuda1998,Yamaya1998} and  $^{44}$Ti \cite{Michel1986A,Michel1986,Michel1988,Yamaya1996,Yamaya1998,Ohkubo1998B,Michel1998}. Secondly,  at  the same time,
it is clear that  the NCM representation of Eq.(\ref{NCM2}) of the GCM wave function
  shows that the $\alpha$ cluster structure   simultaneously has a condensate nature   since the 
   $\alpha$ clusters are   trapped in the  $0s$ state of the harmonic oscillator potential.
 
 %Fig1
\begin{figure}[t] 
\begin{center}
\includegraphics[width=14.0cm]{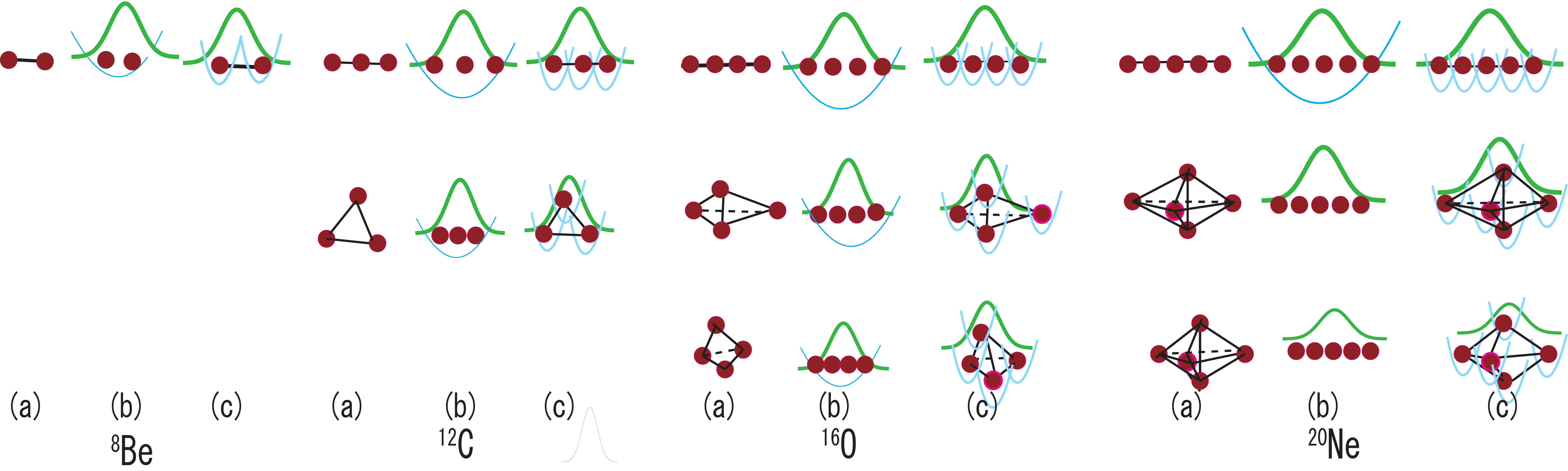}
\end{center}
\caption{ Illustrative   figures  of  crystallinity,  condensation and supersolidity of the  $\alpha$ clusters  (filled circles) in $^8$Be, $^{12}$C, $^{16}$O and $^{20}$Ne.   As the  excitation energy  increases
  vertically, 
  the structure change occurs.  In  each nucleus   (a)  crystallinity, (b)  condensation associated with a coherent wave    and  (c) supersolidity with both  crystallinity and  coherent wave of the $\alpha$ clusters
   are shown. The original Ikeda diagram based on crystallinity picture corresponds to (a) in each nucleus. In (b) of each nucleus $\alpha$ clusters are sitting in the $0s$ state of the 
    harmonic oscillator potential with a  coherent   wave (broad curve).  In (c) of each nucleus the $\alpha$ clusters are sitting in the $0s$ state of the distinct harmonic oscillator potentials separated due to the Pauli repulsion associated with a  coherent   wave (broad curve).
 }
\label{fig1}
\end{figure}
 
       Illustrative pictures based on  the above geometrical structure and the condensate structure  of the   $\alpha$ clusters in $^8$Be,  $^{12}$C, $^{16}$O and $^{20}$Ne are displayed 
         in (a)  and (b) of each nucleus in Fig.~1.   
 The pictures     (a) correspond to the Ikeda diagram  \cite{Ikeda1968,Horiuchi1972}, which  has been a 
 useful guide  
     of   cluster structure study  in nuclei 
      for more than half a century.
The pictures (b) represent  the wave aspect of the $\alpha$ cluster structure due to the condensation.
Thus it is clear that 
 the GCM wave function $\Psi_{n\alpha}^{GCM}$ has  the seemingly opposing  above two aspects, particle nature and coherent  wave nature since $\Psi_{n\alpha}^{GCM}=\Psi_{n\alpha}^{NCM}$ in Eq.(\ref{GCM=NCM-Nalpha-2}).
     The exclusive two pictures, the duality of crystallinity  (localization) and condensate coherent wave  (nonlocalization), can be   reconciled in  the unified  pictures displayed  in (c) of each nucleus in Fig.~1  where the $\alpha$ clusters   sitting in the $0s$ state of  the distinct potentials due to the Pauli  repulsion   between the $\alpha$ clusters \cite{Tamagaki1968} form a coherent wave.
The duality evokes Landau's two-fluid picture of superfluidity of He II   \cite{Brink2005}.
 We  divide the total density  $\rho^{GCM}$ into the two components,
  $\rho^{GCM}$=$\rho^{GCM}_s$+$\rho^{GCM}_n$
where $\rho_s^{GCM}$ is the superfluid density distribution of {the condensate}, which corresponds to (b)  of each nucleus in Fig.~1, and the normal density component $\rho^{GCM}_n$ is defined in the equation.
 It should be noted that $\rho_s^{NCM}$= $\rho_s^{GCM}$, $\rho_n^{NCM}$= $\rho_n^{GCM}$ and  $\rho^{NCM}$= $\rho^{GCM}$ since  $\Psi^{GCM}$=$\Phi^{NCM}$.
 $\rho^{GCM}_s$ is considered to correspond to the superfluid density $\rho^{SCM}_s$ of the superfluid cluster model (SCM) based on  effective field theory, in which the order parameter is embedded by  rigorously treating the Nambu-Goldstone  mode  due  the spontaneous symmetry breaking (SSB) of the global phase 
 \cite{Ohkubo2020}.
The geometrical localization, degree of clustering, is characterized by the order parameter $R$ and the condensation is characterized  by the order parameter,
 superfluid density $\rho_s$.

       For $^8$Be,  in Fig.~1(c),
 the     two $\alpha$ clusters, which can penetrate by quantum tunneling 
 \cite{Gamow1928,Gurney1928,Gurney1929,Tanizaki2014}
   the intercluster barrier due to the Pauli principle,  are  trapped in each  $0s$ state of  the two local minima of  the 
   double-well potential to form  a coherent wave. The  $\alpha$ cluster structure  has the duality of crystallinity and coherent wave due to condensation.  In the case that  the potential has three, four,  $\cdots$, $n$ local minima, it is clear that the three, four,  $\cdots$, $n$ $\alpha$ linear chain  cluster structure  has the duality of crystallinity and condensation. Because of  $\Psi_{n\alpha}^{GCM}=\Psi_{n\alpha}^{NCM}$ in Eq.(\ref{GCM=NCM-Nalpha-2}), whatever the geometrical configuration, number of $\alpha$ clusters,  and degree of clustering,  the GCM $\alpha$ cluster wave function has the duality.
 
In more detail for each nucleus, in   $^{12}$C, 
the  dilute gas-like BEC Hoyle state with  three $\alpha$ clusters  appears near the $\alpha$ threshold.  The Hoyle state and the BEC excited states built on it 
were shown to be reproduced well by  the recent SCM calculations \cite{Nakamura2016,Katsuragi2018,Nakamura2018}. 
 For $^{16}$O, from the geometrical cluster viewpoint, the ground state with  the 
    four $\alpha$ clusters at the vertices of  the tetrahedron \cite{Dennison1954}, which has been recently revisited   in Refs. \cite{Bijker2014,Halcrow2017,Halcrow2019,Halcrow2020},  makes a structure change  to the  $\alpha$+ $^{12}$C ($0_1^+$}) cluster structure near the $\alpha$ threshold as revealed in Refs. \cite{Suzuki1976, Suzuki1976B}
 and   to the loosely  coupled well-developed  four $\alpha$ cluster states  with the dilute gas-like $\alpha$+$^{12}$C($0_2^+$)   cluster structure near the   four $\alpha$ threshold \cite{Ohkubo2010}.   The four $\alpha$ linear chain structure is considered to appear at much higher energies above the four $\alpha$ threshold energy \cite{Ichikawa2011}.
 The above  structure change is consistent with 
    NCM calculations \cite{Funaki2018}   and the BEC four $\alpha$ cluster  calculations  using the SCM \cite{Takahashi2020}.
  For  $^{20}$Ne, the ground state with   five $\alpha$ clusters at the vertices of a trigonal bipyramid \cite{Bouten1962,Brink1968,Brink1970,Nemoto1975,Bijker2020}     makes a structure change  near the $\alpha$ threshold as the excitation energy increases
 \cite{Fujiwara1980}
   and near the five $\alpha$ threshold energy   well-developed   gas-like $\alpha$ cluster BEC superfluid states \cite{Katsuragi2018} are expected to appear as observed in  recent experiments  \cite{Swarz2015,Adachi2020} before the five $\alpha$ linear chain structure at  higher energies.    
Fig.~1 may be extended to 
the  Ikeda diagram in  medium-weight and heavy  nuclei \cite{Ohkubo1998}.

It should be noted that both the NCM  and  the GCM wave functions contain the two aspects of crystallinity  and  condensation.
In other words,   
the nonlocalized wave function does not fully correspond to  Fig.~1(b) of each nucleus  as the localized cluster GCM  wave function does not fully correspond to Fig.~1(a) of each nucleus. 
To what extent the wave function contains the condensate component illustrated in Fig.~1(b)   depends on the degree of clustering of the cluster state. 
  The unified pictures 
   in Fig.~1(c)
evoke  an optical lattice in cold atom physics \cite{Morsch2006,Bloch2008,Yamamoto2013,Leonard2017,Li2017,Tanzi2019,Bottcher2019,Chomaz2019,Tanzi2019A,Natale2019,Guo2019}.
 The $\alpha$ cluster structure with    crystallinity  and condensation, i.e. supersolidity, 
  is a supersolid. 
A supersolid  has been searched for  in recent decades in He II \cite{Andreev1969,Chester1970,Leggett1970,Matsuda1970,Boninsegni2012} and 
has been observed very recently in an optical lattice \cite{Leonard2017,Li2017,Tanzi2019,Bottcher2019,Chomaz2019,Tanzi2019A,Natale2019,Guo2019}.
The structure change in Fig.~1 from the ground state with both  solidity and superfluidity of  $\rho_s$  to the dilute gas-like BEC state near the three and four $\alpha$ threshold is considered to be a phase transition from a supersolid to a superfluid.

% two-fluid picture   
What is the evidence for the  supersolidity? 
The direct evidence of  supersolidity is the observation of 
  a Nambu-Goldstone mode \cite{Nambu1960,Goldstone1960,Nambu1961} due to SSB of the global phase, which  was observed very recently for an optical lattice supersolid \cite{Tanzi2019A,Natale2019,Guo2019}. 
  Since the superfluid density  $\rho_s$ is the order parameter of the SSB of the global phase  \cite{Ohkubo2020}, the existence of $\rho_s$$\ne0$ in the GCM $\alpha$ cluster wave function of the ground state due to the duality  accompanies  the Nambu-Goldstone mode states, which are to be  very low-lying collective states and  difficult to explain in the shell model.  
   This logic is  same as  the emergence of  rotational band states in deformed nuclei, for which  the  order parameter,  deformation parameter, $\delta\ne0$,  caused  by SSB of  rotational invariance due to a quadrupole boson condensation in the ground state \cite{Ring1980}.
  It is known  that the very low-lying intruder  collective $0^+$  states  appear 
    systematically near the $\alpha$ threshold energy in light  and medium-weight nuclei  such as the  mysterious  $0_2^+$
     states in  $^{16}$O
    and 
        in  $^{40}$Ca, which are analog of the intruder $0_2^+$ state in $^{12}$C.
 The  appearance of  such   intruder collective states at a very  low excitation energy near the $\alpha$ threshold, which has been understood by the empirical threshold rule of  the Ikeda diagram \cite{Ikeda1968,Horiuchi1972}, is considered to be  understood from the  viewpoint the Nambu-Goldstone mode due to SSB of the global phase of the $\alpha$ cluster structure as       discussed for  $^{12}$C   in Ref. \cite{Ohkubo2020}.

Finally we mention  the importance of the Pauli principle for the duality of geometrical localization and nonlocalization due to condensation  of $\alpha$ cluster structure.  The geometrical localization of the  $\alpha$ clusters has been known to be  caused by the Pauli principle \cite{Tamagaki1969,Ohkubo2016}. 
In (c) of each nucleus in Fig.~1, the coherent wave    of the  $\alpha$ cluster structure  
 is the consequence of  the  geometrical  localization.
Thus the Pauli principle   has  the dual role of causing  the 
  geometrical clustering and condensation.
 In this sense the origin of the superfluidity of  $\alpha$ cluster structure is  different from that of  the BCS superfluidity in heavy nuclei and cold atoms.
 
 \par%\section{summary}
To summarize, we have shown that the   Brink $\alpha$ cluster model  in the 
generator coordinate method 
 with crystallinity   based on the geometrical picture is mathematically equivalent to the nonlocalized cluster model  based on the 
  condensation of $\alpha$ clusters. Thus the apparently opposing nonlocalized cluster model  is reconciled  with the traditional geometrical localized cluster models such as the Brink cluster model in the generator coordinate method, the resonating group method, the orthogonality condition model,  and  the local potential model, which has been  powerful in understanding    cluster structure in nuclei   intuitively and quantitatively.
 The equivalence is a manifestation of the  duality of the crystallinity and 
  condensation,
particle nature  and  wave nature, of the geometric cluster structure. 
 The $\alpha$ cluster structure is understood to have crystallinity and condensation simultaneously,   that is,  supersolidity. The Pauli principle causes  the duality.
 The evidence of 
 supersolidity of  $\alpha$ cluster structure  is the emergence of the Nambu-Goldstone mode due to the SSB of the global phase, i.e. emergence of a collective motion at very low excitation energy. The emergence of the $\alpha$ cluster states
at very low excitation
   near the $\alpha$ threshold such as the Hoyle state $0^+_2$ of $^{12}$C and the mysterious  $0^+_2$ state of $^{16}$O are considered to be  such a member state of  the  Nambu-Goldstone mode.

\ack
The author thanks the Yukawa Institute for Theoretical Physics, Kyoto University   for   the hospitality extended  during  a stay in  2019.

\end{document}